\documentclass[twocolumn]{article}

\usepackage{epsfig}

\title{Linear Order Matrix Inversion Method\\with Help from Quantum Searching Algorithm}
\author{Atsushi Miyauchi \medskip
\\{\it Research Organization for Information Science and Technology (RIST)}
\\{\it 2-2-54 Nakameguro, Meguro-ku, Tokyo 153-0061, Japan}
\\{\tt miyauchi@tokyo.rist.or.jp}}
\date{\null}
\newtheorem{lemma}{Lemma}

\begin{document}

\twocolumn[
\flushright{ERATO Workshop on\\{\bf Quantum Information Science}\\{\sl University of Tokyo, 6-8 Sep. 2001}}
\maketitle
\vspace{-10mm}
\begin{center}
{\textbf{\bf Abstract}\\\medskip}
\newlength{\abswidth}
\setlength{\abswidth}{\linewidth}
\addtolength{\abswidth}{-3cm}
\begin{minipage}{\abswidth}
{\small Presented here is a matrix inversion method utilizing quantum searching algorithm. In this method, huge Hilbert space as a whole spanned by myriad of eigen states is searched and evaluated efficiently by sequential reduction in dimension one by one.
Total iteration steps required for search are proportional to the number of unknown variables. Our method could solve very large linear equations with sufficiently high probability faster than any existing classical algorithms, which roughly depends on the cube of unknown variables.}
\end{minipage}\\
\vspace{4mm}
\vspace{5mm}
\end{center}
]
%
\section{Introduction}
Since Grover\cite{Grover} has first invented his quantum searching algorithm quadratically faster than classical counterparts, several applications of the algorithm have been considered, for example, finding the minimum\cite{Durr}, estimating the median\cite{Grover2}, quantum counting\cite{Brassard}, collision problem\cite{Brassard2}, undirected graph connectivity\cite{Watrous} and protein sequence comparison\cite{Hollenberg}. 
On the other hand, this searching algorithm is thought to be unsuitable for such massively number crunching computer simulations like aerodynamics, car crush, electric circuitry, weather prediction, molecular biology, finance and so on. 
Frequently, large part of those simulations are devoted to solve linear equations, which are mathematically written in matrix form. In this paper, we present a method to solve matrix equation efficiently on quantum computer by utilizing quantum searching algorithm.
In our method, huge Hilbert space as a whole spanned by myriad of eigen states is searched and evaluated efficiently by sequential reduction in dimension one by one.

In the next section, we describe the matrix inversion method both naive and improved in detail. The third section estimates the expected computational performance numerically. The final section concludes this paper.
%
\section{Quantum matrix inversion}
Any system of linear equations can be written as single matrix equation as following
\[
	Ax+b=0 \nonumber\,,
\]
where $A$ denotes matrix with $n$-columns and $m$-rows, and $b$ denotes vector with $n$-rows.
For simplicity, we assume that the system is neither overdeterminate nor underdeterminate. This assumption imposes $n=m$ and $\det(A) \neq 0$ on matrix $A$ and guarantees unique solution. The problem to be solved is to locate $x$ satisfying above equation for given $A$ and $b$. This kind of problem has been investigated for many years and the baseline algorithm was discovered by great German mathematician, Carl Friedrich Gauss, in nineteenth century. Unfortunately, his algorithm sometimes falls in instability and inaccuracy for a large matrix on digital computers mainly because of accumulation and amplification of truncation error. Although many improvements, enhancements and innovation have been persued on classical algorithm, comutational workload propotional to the cube of unknown variables is still required. On the other hand, quantum algorithm utilizes novel characteristics of quantum mechanics known as quantum parallelism to accelerate its computation dramatically. Especially, Grover's\cite{Grover} quantum searching algorithm is quite versatile to use and already applied in several problems. In the following subsections, we propose the matrix inversion method based on his algorithm to obtain further acceleration from classical counterparts.
%
\subsection{Naive implementation}
In this subsection, we describe the straightforward implementation of Grover's algorithm on matrix inversion. For the first time, we prepare $3n$ null registers $|{\mathbf 0}\rangle$
\[
	|{\mathbf 0}\rangle = \underbrace{|0\rangle|0\rangle \cdots |0\rangle}_{3n} \,.
\]
Operating Walsh-Hadamard gates on the first $n$ registers provides uniformly superposed initial state $|\psi(0)\rangle$, which forms discrete $n$-dimensional searching space as following
\begin{eqnarray}
	|\psi(0)\rangle 
	&=& H^{\otimes n} \otimes I^{\otimes 2n}|{\mathbf 0}\rangle \nonumber\\
	&=& \!\!\frac{1}{\sqrt{M^n}} \sum_{x_1=1}^M\cdots\sum_{x_n=1}^M |x_1\rangle\cdots|x_n\rangle \underbrace{|0\rangle \cdots |0\rangle}_{2n} \,, \nonumber
\end{eqnarray} 
where $H$ and $I$ denote Walsh-Hadamard gate and identity operator, and $M$ denotes the number of numerical points in each dimension.
Therefore $|\psi(0)\rangle$ forms $M^n$ dimensional Hilbert space.
Succeedingly, we multiply the first row of matrix $A$ with $x$. According to Vedral {\it et. al.} \cite{Vedral}, there exist unitary operators $U_{11}, U_{12}, \cdots, U_{1n}$ for each matrix elements $a_{11}, a_{12}\cdots a_{1n}$ which map $x_1, x_2\cdots x_n$ to $a_{11}x_1, a_{12}x_2\cdots a_{1n}x_n$ such that
\[
	U_{ij}|x_j\rangle|0\rangle = |x_j\rangle|a_{ij}x_j\rangle \,.
\]
After exertion of these operators, we obtain
\begin{eqnarray}
	|\psi(1)\rangle 
	&=& U_{11}\otimes U_{12}\otimes\cdots\otimes U_{1n}\otimes I^{\otimes n} |\psi(0)\rangle \nonumber\\
	&=& \frac{1}{\sqrt{M^n}} \sum_{\mbox{\boldmath$x$}} |x_1\rangle\cdots|x_n\rangle \nonumber\\
	&& \qquad |a_{11}x_1\rangle\cdots|a_{1n}x_n\rangle \underbrace{|0\rangle \cdots |0\rangle}_{n} \,, \nonumber
\end{eqnarray}
where we introduced an abbreviation $\mbox{\boldmath$x$}$, which denotes a set consists of all $n$ dimensional numerical points, that is, $\mbox{\boldmath$x$} = \{ x_1, x_2, \cdots, x_n : 1\le x_i\le M\,\, \mbox{for each \it i}\}$.
Then, we take summation of the middle registers to obtain $f_1$ defined as
\[
	f_1 = \sum_{j=1}^n a_{1j} x_j + b_1 \,,
\]
where $b_1$ denotes the first element of vector $b$.
Since addition also can be realized by some unitary operators\cite{Vedral}, we can calculate above equation through quantum gates. 
Here we obtain following state
\begin{eqnarray}
	|\psi(2)\rangle 
	&=& \frac{1}{\sqrt{M^n}} \sum_{\mbox{\boldmath$x$}} |x_1\rangle\cdots|x_n\rangle \nonumber\\
	&& \qquad |a_{11}x_1\rangle\cdots|a_{1n}x_n\rangle |f_1\rangle \underbrace{|0\rangle \cdots |0\rangle}_{n-1} \,. \nonumber
\end{eqnarray}
Obviously, the middle registers are useless for the rest of calculation, so that we clear up such garbage and recycle them in the next calculation step. Garbage erasure is realized by backward operation, which has first devised in connection with reversible computer by Bennett\cite{Bennett}. After garbage erasure we have
\begin{eqnarray}
	|\psi(3)\rangle 
	&=& \frac{1}{\sqrt{M^n}} \sum_{\mbox{\boldmath$x$}} |x_1\rangle\cdots|x_n\rangle \nonumber\\
	&& \qquad  \underbrace{|0\rangle \cdots |0\rangle}_{n} |f_1\rangle \underbrace{|0\rangle \cdots |0\rangle}_{n-1} \,. \nonumber
\end{eqnarray} 
Repeating above operations for the rest of $n-1$ rows, eventually we obtain
\begin{eqnarray}
	|\psi(3n)\rangle 
	&=& \frac{1}{\sqrt{M^n}} \sum_{\mbox{\boldmath$x$}} |x_1\rangle\cdots|x_n\rangle \nonumber\\
	&& \qquad  \underbrace{|0\rangle \cdots |0\rangle}_{n} |f_1\rangle\cdots|f_n\rangle \nonumber \,.
\end{eqnarray} 
This finishes preparation. From here, Grover's quantum searching algorithm is invoked. 
To utilize his algorithm, we must identify oracle $C$ which determine whether the argument satisfies given constraints. Oracle must have a property like that
\[
	C = \left\{ \begin{array}{rl} 1 & \mbox{if $\sum^n_{i=1}|f_i| = 0$} \\ 0 & \mbox{otherwise} \end{array} \right. .
\]
However, particular construction of above oracle is left for future work currently.
Meanwhile, Grover showed that only unique solution survives after a number of iterations of state rotation.
Necessary iteration count has estimated exactly by Boyer {\it et al. }(Hereafter we refer it as BBHT)\cite{Boyer} as close to $\frac{\pi}{4}\sqrt N$, where $N$ denotes the total number of candidates in search. They also found the failure probability as $1/N$.
In case of naive implementation described here, $N$ equals $M^n$. Consequently, computational steps required for search amount to roughly $M^{\frac{n}{2}}$. While failure probability is sufficiently low, the exponential dependency on $n$ forces computational steps to explode, and therefore
 makes naive implementation definitely impractical for especially large-scale matrices.
%
\subsection{Dimensional reduction}
As you see in the previous subsection, searching such a huge Hilbert space as a whole would be a desperate effort even if quantum computer were available.
In this subsection, we employ {\it "a box in a box"} strategy which, in practice, divide the searching space into a sequence of lower dimensional subspaces one by one. Basically procedures required here are similar to the previous ones. However, we put BBHT algorithm instead of Grover's algorithm immediately after the calculation of $|\psi(3)\rangle$, since BBHT can find out multiple solutions. In each iteration step, we use oracle $C'$ such as
\[
	C' = \left\{ \begin{array}{rl} 1 & \mbox{if $f_1 = 0$} \\ 0 & \mbox{otherwise} \end{array} \right. ,
\]
instead of $C$.
According to BBHT, multiple solutions are obtained after approximately $\frac{\pi}{4}\sqrt{N/t}$ iterations with failure probability $t/N$, where t denotes the number of solutions.
Provided that the following inequality similar to diagonally dominance holds for any $i$
\[
	2 \max_j|a_{ij}| \le \sum_j |a_{ij}| + \frac{|b_i|}{M} \,,
\]
then $t$ always forms $n-1$ dimensional complete, {\it i.e.} unclipped, intersection.
In such cases, we can assure that $N=M^n$ and $t=M^{n-1}$. Therefore, iteration count and failure probability in this stage can be estimated as approximately $\frac{\pi}{4}\sqrt M$ and $1/M$ respectively. After the search in the first dimension, we obtain the following state
\begin{eqnarray}
	|\psi(4)\rangle 
	&=& \frac{1}{\sqrt{M^{n-1}}} \sum_{\mbox{\boldmath$x'$}} |x_1\rangle\cdots|x_n\rangle \nonumber\\
	&& \qquad  \underbrace{|0\rangle \cdots |0\rangle}_{n} |f_1\rangle \underbrace{|0\rangle \cdots |0\rangle}_{n-1} \,, \nonumber
\end{eqnarray} 
where $\mbox{\boldmath$x'$}$ denotes a subset of $\mbox{\boldmath$x$}$, that is, $\mbox{\boldmath$x'$} = \{ x_1, x_2, \cdots, x_n : f_1=0\,;\,\, 1 \le x_i \le M \,\, \mbox{for each \it i} \}$.
In other words, only $n-1$ dimensional subspace survives through the first searching stage.
Next, we multiply the second row of matrix $A$ with surviving superposed state $\mbox{\boldmath$x'$}$ as just like the way we did previously. Succeedingly, summation and garbage erasure follow. Then we put BBHT again. Oracle to be used here is
\[
	C'' = \left\{ \begin{array}{rl} 1 & \mbox{if $f_2 = 0$} \\ 0 & \mbox{otherwise} \end{array} \right. ,
\]
instead of $C'$.
This time, volume of searching space is $M^{n-1}$ and number of solutions are $M^{n-2}$. Thus, necessary iteration count and failure probability are the same as before, 
that is, $\frac{\pi}{4}\sqrt M$ and $1/M$ respectively. After the search in the second dimension, the state is
\begin{eqnarray}
	|\psi(5)\rangle 
	&=& \frac{1}{\sqrt{M^{n-2}}} \sum_{\mbox{\boldmath$x''$}} |x_1\rangle\cdots|x_n\rangle \nonumber\\
	&& \qquad  \underbrace{|0\rangle \cdots |0\rangle}_{n} |f_1\rangle |f_2\rangle \underbrace{|0\rangle \cdots |0\rangle}_{n-2} \,, \nonumber
\end{eqnarray} 
where $\mbox{\boldmath$x''$}$ denotes a subset of $\mbox{\boldmath$x'$}$, that is, $\mbox{\boldmath$x'$} = \{ x_1, x_2, \cdots, x_n : f_1=0\,;\,\,f_2=0\,;\,\, 1 \le x_i \le M \,\, \mbox{for each \it i} \}$.
After repetition of this procedure $n-2$ times for the rest rows of matrix $A$, searching subspace reduces to zero dimension. In this way, finally we get to the unique solution
\begin{eqnarray}
	|\psi_{final}\rangle 
	&=& |x_1^{solution}\rangle\cdots|x_n^{solution}\rangle \nonumber\\
	&& \qquad  \underbrace{|0\rangle \cdots |0\rangle}_{n} |f_1\rangle \cdots |f_n\rangle \,. \nonumber
\end{eqnarray} 
Here, we measure the registers to read out solutions.
We can readily see that the summation of all iteration count amounts to $\frac{\pi}{4}n\sqrt M$.
To get to the true solution, all of $n$ searches should be succeeded. Thus, success probability for the total procedure equals to multiplication of success probability in each search, that is, $(1-1/M)^n$.
Obviously, iteration count scaling with $n$ shows exponential acceleration from naive implementation. This fact shows that dimensional reduction technique described here works well for the large matrix inversion.
%
\section{Performance}
In this section, we will investigate performance issues in detail.
As shown in previous section, number of steps required in searching algorithm is $\frac{\pi}{4}n\sqrt M$.
In addition to this, $n^2$ steps are also needed to calculate inner product.
Regretfully, we don't have any idea to make them decrease for the time being.
Meanwhile, current digital computer usually assigns four bytes data({\it i.e.} 32bits), for each real number. Here we also adopt this as typical qubit size of register, that is, $M=2^{32}$.
Now we can readily estimate operational count required for entire computation as $2n(51,471+n)$, in which the factor of two reflects backward operation to erase garbage.
On the other hand, it is well known that classical Gaussian elimination method requires the order of $n^3$ steps\cite{Press}.
From these expressions, we can estimate crossover size from classical to quantum algorithm as roughly $n=321$.
This result means that even relatively small matrix can be accelerated by quantum algorithm.
However, note that the estimation here can not be taken serious for the time being, since operational speed per gate of current digital computer is considerably faster than that of today's infant quantum computer.

Another important issue to be considered besides iteration count is success probability.
If the probability to obtain true solution is poorly low, we should be hopelessly exhausted recalculating so many times.
In previous section, the success probability of dimensional reduction technique is written as $(1-1/M)^n$. To find out the lower bound of this probability, we prove following two lemmas.

\begin{lemma}\label{lem1}
Let $k$ and $n$ be any natural numbers such that $k\le n$, then following inequality holds
\[
	{}_nC_k \le n^k \,.
\]
Proof. 
We prove this lemma by induction. For $k=1$, above inequality obviously holds for any natural number $n$. Provided that the inequality holds for a integer $k$ such that $1<k<n$, LHS term for $k+1$ could be estimated as
\[
	{}_{n}C_{k+1} = \frac{(n-k)}{k+1} {}_nC_k \le  \frac{(n-k)}{k+1} n^k \le n^{k+1} \,.
\]
Notice that $\frac{(n-k)}{k+1}<n$ holds for any positive $n$ and $k$. Consequently, induction principle guarantees above inequality for any $k$. 
\hfill \setlength{\fboxsep}{0pt}\setlength{\fboxrule}{2pt}\fbox{\tiny }
\end{lemma}

\begin{lemma}\label{lem2}
Let $p$ be a real number such that $0\le p\le 1$ and $n$ be a any natural number such that $np<1$,  then following inequality holds
\[
	\left(1-p\right)^n > \frac{1-2np}{1-np} \,.
\]
Proof.
Using binary term expansion,
\begin{eqnarray*}
	\left(1-p\right)^n &=& \sum^n_{k=0} {}_nC_k(-p)^k \\
	&>& 1-\sum^n_{k=1} {}_nC_k \,p^k \,.
\end{eqnarray*}
Substituting the result of lemma \ref{lem1}, the estimation continues like
\begin{eqnarray*}
	\qquad\qquad
	&>& 1-\sum^n_{k=1} (np)^k \\
	&=& 1-\frac{np-(np)^{n+1}}{1-np} \\
	&>& \frac{1-2np}{1-np} \,.
\end{eqnarray*}
Here completes the proof. \hfill 
\hfill \setlength{\fboxsep}{0pt}\setlength{\fboxrule}{2pt}\fbox{\tiny }
\end{lemma}

Substituting $p=1/M$ in lemma \ref{lem2} gives
\[
	\left(1-\frac{1}{M}\right)^n \ge \frac{1-2n/M}{1-n/M} \,.
\]
Now, we estimate this lower bound of success probability numerically.
Substituting $M=2^{32}$ as typical number described before, we can make sure that success probability is no less than 93\% for $n\le 2^{28} \simeq 10^9$ and that the smaller $n$, the higher success probability for fixed $M$.

In this section, it is shown that quantum matrix inversion enhanced by dimensional reduction technique could solve matrix equation with sufficiently high probability within linear time, even if its problem size were either relatively small or considerably large.
%
\section{Conclusion}
In this paper, we proposed a matrix inversion method utilizing quantum searching algorithm.
The heart of our method is dimensional reduction techneque, which is introduced to accelerate searching procedure.
This technique enables us to obtain exponential speed-up over naive implementation and sufficiently high probability of success.
Required number of iteration steps linearly depends on the problem size, while quadratic steps are necessary for arithmetic calculations.
Proposed method achieves acceleration over classical algorithms by an order of $n$.
These results might open a possibility of quantum computer in future industrial use.
Actually, our proposal is nothing but a primary desktop calculation and we need further investigation.
For example, to identify particular gate configuration, to estimate performance stringently through emulation of gate operation and to understand susceptibility to error caused by truncation and decoherence are left as future works.
%
\bibliographystyle{plain}

%
\end{document}